\documentclass{aa}

\usepackage{graphicx}
\usepackage{amsmath}
\usepackage{ulem}
\usepackage{amssymb}
\usepackage{txfonts}
\usepackage{xparse,xcolor}

\begin{document}

\title{Modelling hystereses observed during dwarf-nova outbursts}

\author{J.-M. Hameury\inst{1}
    \and
          C. Knigge\inst{2}
    \and
          J.-P. Lasota\inst{3,4}
    \and
    	  F.-J. Hambsch\inst{5,6,7}
    \and
    	  R. James\inst{7}}

\institute{
     Observatoire Astronomique de Strasbourg, Université de Strasbourg, CNRS UMR 7550, 67000 Strasbourg, France\\
   		    \email{jean-marie.hameury@astro.unistra.fr}
   \and
     School of Physics and Astronomy, University of Southampton, Southampton SO17 1BJ, UK,
   \and
     Institut d'Astrophysique de Paris, CNRS et Sorbonne Universit\'es, UPMC Paris~06, UMR 7095, 98bis Bd Arago, 75014 Paris, France
\and
     Nicolaus Copernicus Astronomical Center, Polish Academy of Sciences, Bartycka 18, 00-716 Warsaw, Poland
\and
Vereniging Voor Sterrenkunde (VVS), Oostmeers 122 C, 8000 Brugge, Belgium
\and
Bundesdeutsche Arbeitsgemeinschaft für Veränderliche Sterne e.V. (BAV), Munsterdamm 90, D-12169 Berlin, Germany
\and
American Association of Variable Star Observers, 49 Bay State Road, Cambridge, MA 02138, USA
}

   \date{}


  \abstract
   {Although the disc instability model is widely accepted as the explanation for dwarf nova outbursts, it is still necessary to confront its predictions to observations because much of the constraints on angular momentum transport in accretion discs  are derived from the application of this model to real systems.}
   {We test the predictions of the model concerning the multicolour time evolution of outbursts for two well--observed systems, SS Cyg and VW Hyi.}
   {We calculate the multicolour evolution of dwarf nova outbursts using the disc instability model and taking into account the contribution from the irradiated secondary, the white dwarf and the hot spot.}
   {Observations definitely show the existence of a hysteresis in the optical colour-magnitude diagram during the evolution of dwarf nova outbursts. We find that the disc instability model naturally explains the existence and the orientation of this hysteresis. For the specific cases of SS Cyg and VW Hyi, the colour and magnitude ranges covered during the evolution of the system are in reasonable agreement with observations. However, the observed colours are bluer than observed near the peak of the outbursts -- as in steady systems, and the amplitude of the hysteresis cycle is smaller than observed. The predicted colours significantly depend on the assumptions made for calculating the disc spectrum during rise, and on the magnitude of the secondary irradiation for the decaying part of the outburst.}
   {Improvements of the spectral disc models are strongly needed if one wishes to address the system evolution in the UV.}


   \maketitle
%

\section{Introduction}
Dwarf novae are a subclass of cataclysmic variables that undergo regular 4--6 mag outbursts lasting from one to few days, with a typical recurrence time of a few weeks \citep[see][for a review]{w03}. It is now widely accepted that these outbursts are caused by a thermal-viscous instability of the accretion disc that occurs when the disc temperature is of the order of the ionisation temperature of hydrogen, and the opacities become strongly temperature-dependent \citep[see][for reviews of the model]{l01,Hameury19}.
                                                                                                                                                            
Despite the fact that the disc instability model (DIM) is based on the assumption that angular momentum transport in the disc can be described by an effective viscosity depending on a single parameter \citep{SS73}, the DIM has been quite successful in explaining the overall properties of dwarf nova outbursts. Yet we know now that the assumption that energy dissipation is local and due to the transport of angular momentum, and that it can be described by the Shakura-Sunyaev viscosity is at best very crude. The magnetorotational instability (MRI) is most probably responsible for angular momentum transport during bright states \citep{bh91}, when the disc is hot and fully ionised, but the situation could be quite different during the quiescence \citep[see, e.g.,][]{Coleman16,Scepi18}. Progress in our understanding of angular momentum transport in accretion discs will come both from theoretical studies of the various physical processes that might be involved, but also from the confrontation of the predictions of the DIM with observations.

The DIM reproduces well the timing properties of dwarf nova outbursts, but the wealth of observations extends well beyond light curves. Eclipse mapping provides the brightness variations along the disc radius \citep{h85}, and  hence the radial dependence of the effective temperature that can be directly compared to models. Observational results do not always seem to agree well with predictions from the DIM: in the case of EX Dra, the quiescent temperatures are too high, and their radial distribution corresponds to that of a steady disc \citep{b16}, in particular in the inner disc. It is unclear if this is a deficiency of the DIM or is due to systematic effects generated by the fact that the inclination of eclipsing systems is high \citep[$85^\circ$ in the case of EX Dra; see][for high inclination effects on the temperature profiles]{Smak94}.  In other systems, with lower inclinations, such as Z Cha \citep{hc85}, the agreement is better. Spectral fitting is not restricted to systems that are almost edge-on, but faces the difficulty that models, even the sophisticated ones that take into account effects such as variations of the vertical component of gravity with altitude, or viscous heat dissipation in the disc atmosphere \citep{h90} do not reproduce well the observed spectra of novalike systems, in particular in the UV; we refer to \citet{np19} for a recent discussion of this problem and \citet{w84a,w84b,klb97,ik97,klw98,pdf07} for an historical perspective. Fitting the details of time-dependent spectra is then unlikely to provide much information on dwarf novae; this conclusion is reinforced by the fact that our understanding of the spectra in the low state is very poor, despite some efforts in the past years \citep{ilh10}.

For these reasons, and despite an early work by \citet{ck87}, confrontations with observations have been limited to colours, and notably to the so-called UV delay at a time when it was believed that the time-lag between the rise in the UV and in the optical was an indicator of the inside-out (the outburst is triggered in the inner part of the disc) or outside-in nature of the dwarf nova eruption. \citet{shl03} showed that this was not the case, because a significant increase of the UV requires that the mass accretion rate at the inner disc edge also increases by a large fraction, which takes longer  than the propagation time of the heating front, occuring on a fraction of the viscous time estimated in the inner parts of the disc.

\begin{figure*}
\includegraphics[width=17cm]{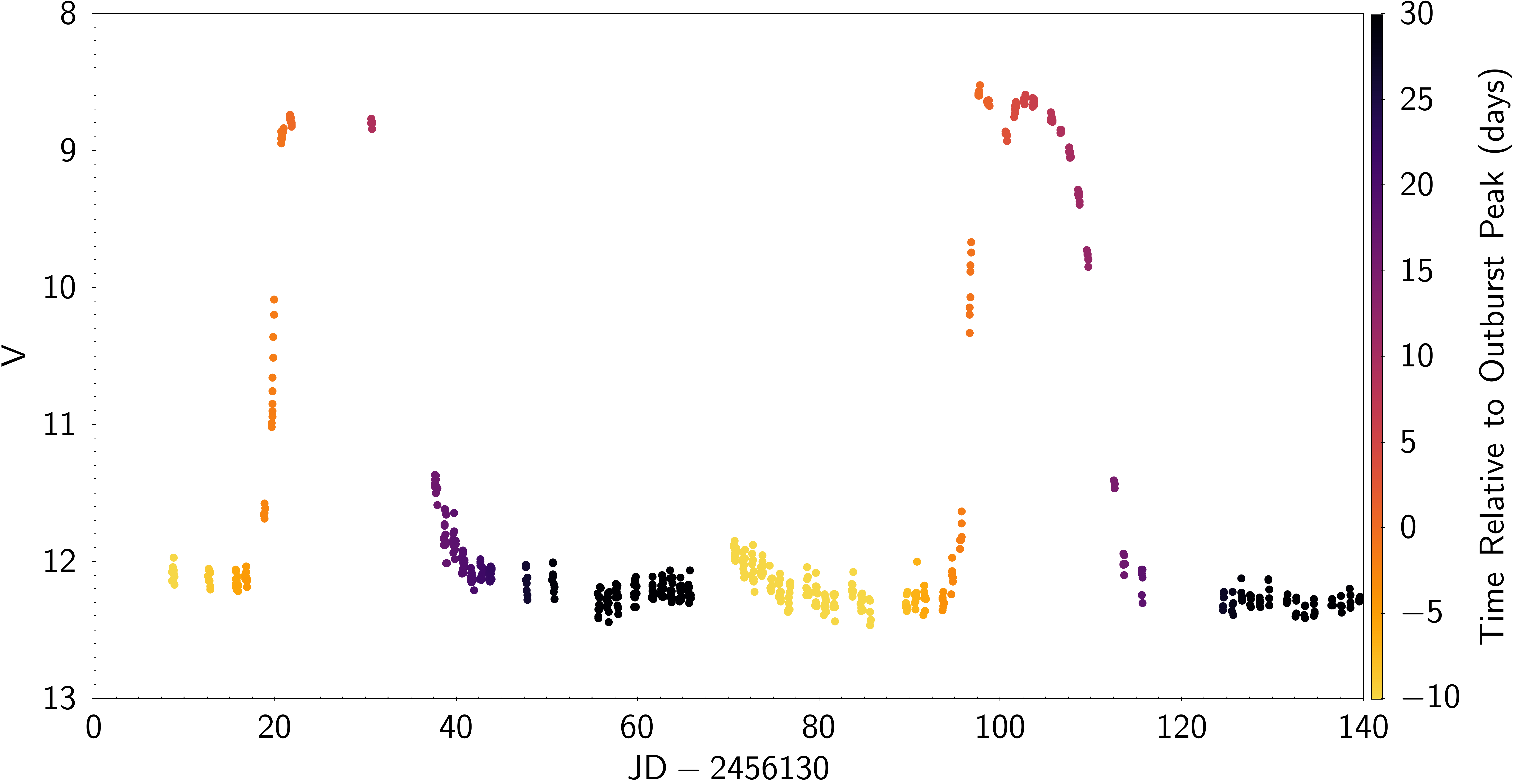}
\caption{SS Cyg light curve from the AAVSO in the V band.}
\label{fig:sscyg}
\end{figure*}

\begin{figure*}
\includegraphics[width=17cm]{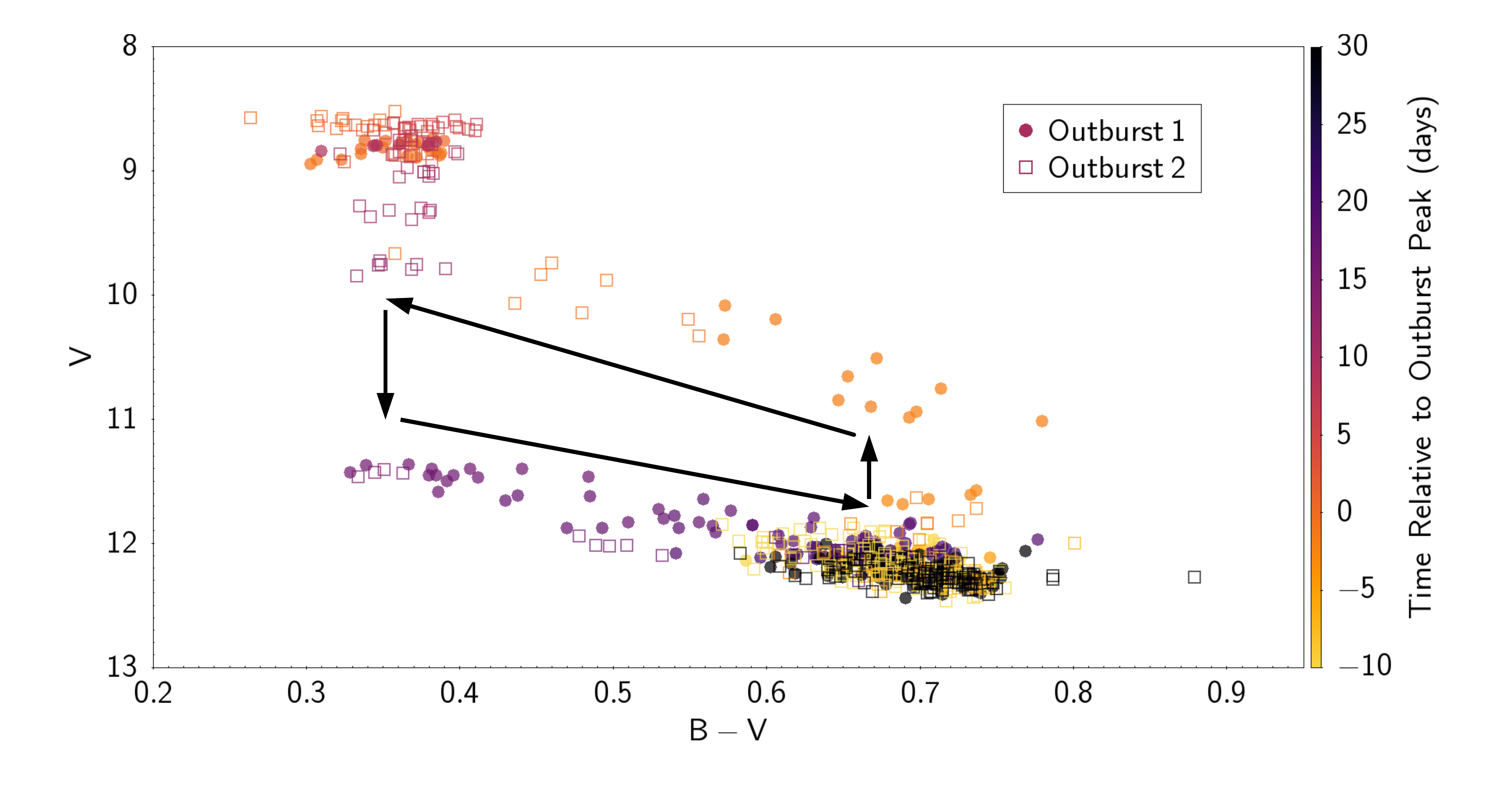}
\caption{Colour variations across two outbursts of SS Cyg. The colour coding is the same as in Fig. \ref{fig:sscyg}, and represents the time. The arrows show the direction of the time evolution in the diagram.}
\label{fig:sscyg_cmd}
\end{figure*}

In this paper, we calculate the colour variations in  $B$ and $V$ that are predicted by the DIM and compare them to observations; we also briefly discuss the situation in the ultraviolet (U band). We improve on the early work by \citet{ck87} by using a more elaborate version of the DIM that solves the full vertical structure of discs, has sufficient spatial resolution to resolve the structure of heat fronts, and allows for variations of the outer disc radius, which has a profound influence on the outburst cycle \citep{hmd98}, in particular on the occurrence of outside-out versus inside-out outbursts. \citet{ck87} assumed that matter was added to the disc at the circularisation radius, that is the radius at which matter leaving the $L_1$ point would form circular orbits while keeping its initial angular momentum; we assume here that it is added at the outer disc edge. We also take into account the contribution of the secondary star and of the hot spot which can be very significant in particular for long orbital period systems. We finally account for the specific parameters of the observed systems. On the other hand, we use a simplified treatment of the disc spectral properties by not taking into account the energy dissipation in the photosphere when estimating the flux emitted at  different spectral  bands.

It has been known for long that, in the colour-colour $(U-B,B-V)$ plane, a dwarf nova such as VW Hyi follows a loop \citep{b80}. This hysteresis type behaviour is similar to the one found when considering variations of the optical flux versus the X-ray to EUV plus X-ray flux in SS Cyg \citep{hlk17}, and, as we shall see later, is accounted for by the same effect: a complete change of the disc's mass--distribution  during outburst. Soft X-ray transients (SXTs) that are the analogues of dwarf novae in low-mass X-ray binaries also exhibit a hysteresis in the hardness-intensity diagram but despite apparent similarities with the one found in dwarf novae \citep{krk08} its origin is different and unrelated to the DIM \citep{hlk17} because for SXTs the viscous time scale in the X-ray emitting regions is much shorter than the outburst duration, and the DIM does not predict any difference between the X-ray properties of the rise and decay of disc outbursts. 

In Sect. \ref{sec:obs}, we present the available observations in the visible for SS Cyg and VW Hyi from the AAVSO and in particular the time evolution in a colour-magnitude diagram that extends \citet{b80} observations to the visible band and are easier to interpret; in Sect. \ref{sec:model} we describe our methods and assumptions for estimating the integrated colours, and we finally present out results together with the comparison with observations both from the AAVSO and from \citet{b80} in Sect. \ref{sec:results}. The objective here is not to fit precisely the observational data, given the uncertainties and approximations inherent to the DIM, but rather to check if the general characteristics of the observations can be accounted for by the DIM.

\begin{figure*}
\includegraphics[width=17cm]{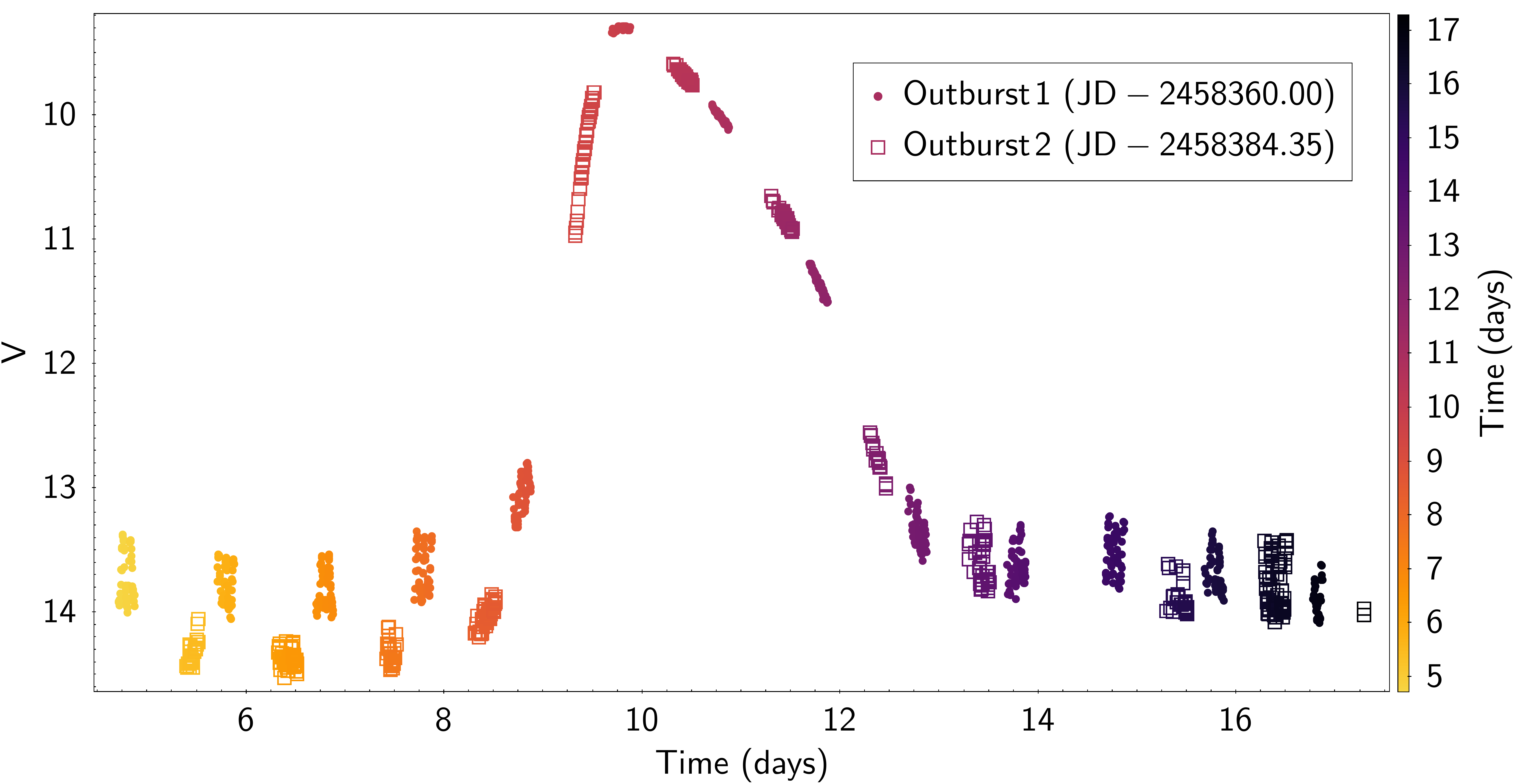}
\caption{VW Hyi light curve from the AAVSO in the V band.}
\label{fig:vwhyi}
\end{figure*}

\begin{figure*}
\includegraphics[width=17cm]{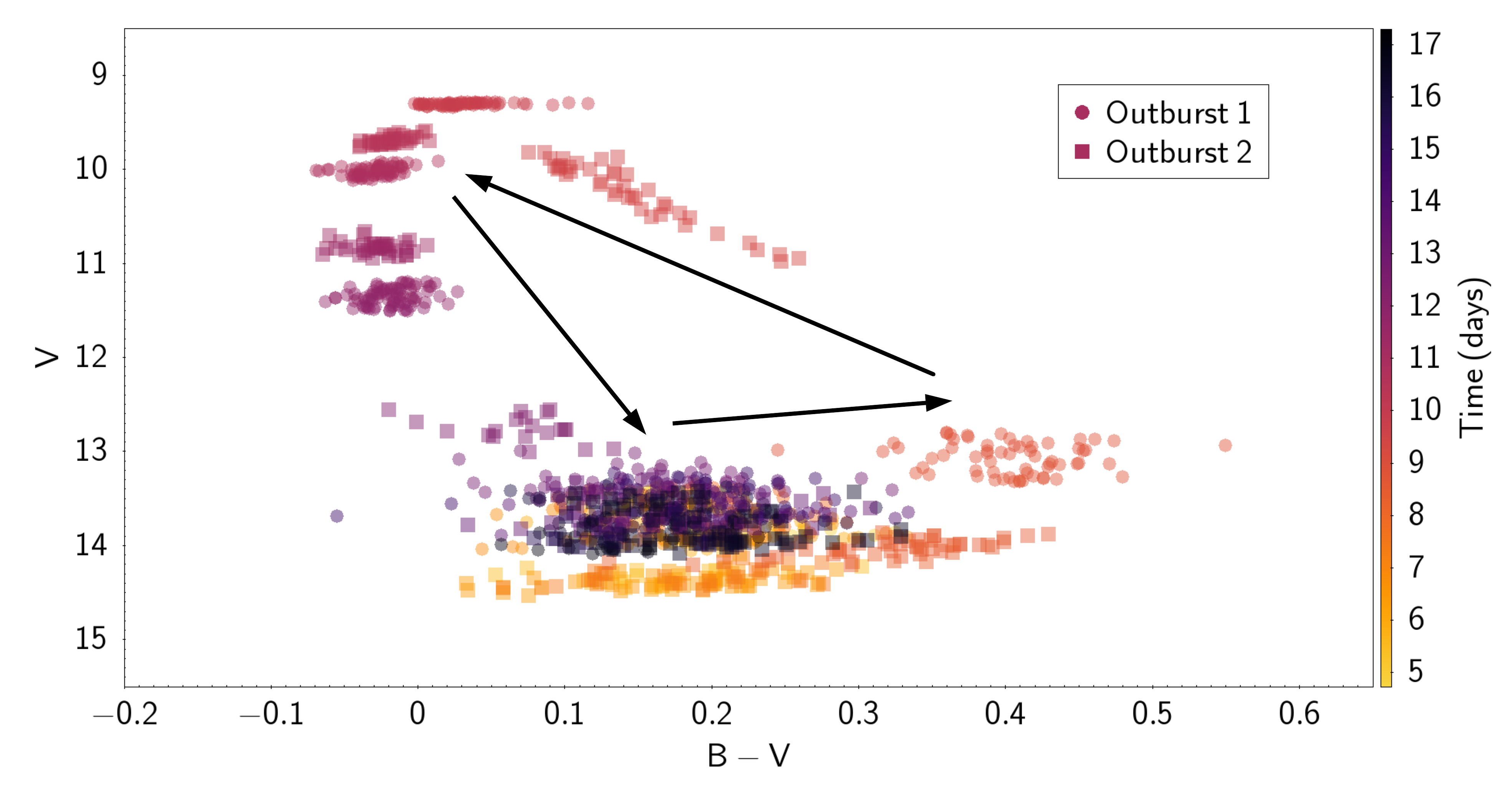}
\caption{Colour variations across an outburst of VW Hyi. The arrows show the direction of the time evolution in the diagram.}
\label{fig:vwhyi1}
\end{figure*}

\section{Observational data \label{sec:obs}}

SS Cyg, with an orbital period of 6.60~h, is one of the best observed dwarf novae, with en extended historical coverage by the AAVSO beginning at the end of the 19th century. We selected two well covered consecutive outburst with available data in the $B$ and $V$ bands (the $B$ magnitude being obtained in most cases 20 min earlier than the $V$ magnitude) obtained by the same observer \citep{jm}, with reasonable error bars. The first outburst was triggered on Aug. 9, 2012, and lasted for slightly more than two weeks; the second one started on Oct. 25, 2012, (i.e. 76 days later) and also lasted for 2 weeks. Figure \ref{fig:sscyg} shows the light curve of these two outbursts. The SS Cyg outburst distribution is bimodal, and consists of long (2 weeks) and bright outbursts and a roughly equal number of short (one week) and slightly fainter ones \citep{bv83}; the outbursts we have selected falls in the first category, with in addition a doubled-peak light curve for the second outburst, which is relatively rare but not exceptional. The observed colour-magnitude diagram is shown in Fig. \ref{fig:sscyg_cmd}. The statistical uncertainties on $B-V$ are typically around 0.03~mag. Systematic errors may be larger, but should be less than 0.1~mag, although there could be some colour dependence in this. The systematics may, however, be different for bright ($V < 11$) and faint ($V > 11$) phases, due to the use of different comparison stars. In quiescence, $B-V$ is approximately 0.7, and SS Cyg becomes bluer in outburst, with a $B-V$ reaching 0.35; this is not unexpected, but what is more remarkable is that the colours during the rise and the decay are significantly different, the rise being redder than the decay.

We then consider the case of VW Hyi, which is also a source well--observed by amateur astronomers, and we have selected two consecutive outbursts from the AAVSO database that occurred on 2018, Sep. 7 and Oct. 2 \citep{hmb}; these are shown in Fig. \ref{fig:vwhyi}. As can be seen, these two outbursts have remarkably similar shapes. VW Hyi has an orbital period of 1.78~h, and is the brightest and the best observed system belonging to the SU UMa subclass that presents, in addition to normal outbursts, brighter and much longer superoutbursts. It is generally believed that superoutbursts are caused by a tidal instability coupled to the thermal-viscous instability \citep{o89}, but the validity of this  model has been questioned by \citet{BH02}, \citet{s09} and \citet{c12} who favoured a model based on irradiation enhanced mass transfer from the secondary \citep{HLW00}. Whatever the mechanism accounting for superoutbursts, the two bursts we have selected belong to the category of normal outbursts, and are caused by the classical thermal-viscous instability.

The colour-magnitude diagram of VW Hyi is shown in Fig. \ref{fig:vwhyi1}. As for SS Cyg, the statistical uncertainties on $B-V$ are typically around 0.03~mag., and the systematic errors less than 0.1~mag., with no difference between bright and faint states due to the use of the same comparison stars. This diagram is very similar to that of SS Cyg, the main difference being that (i) VW Hyi is bluer both in quiescence and in outburst, and (ii) that during the initial rise, the system becomes redder by almost half a magnitude. In both cases, an hysteresis is clearly visible with an amplitude similar to that of SS Cyg. 

These results are, as far as the B and V magnitudes are concerned, comparable to those of \citet{b80}; our colours at a given magnitude are similar to his, and he also finds that during the rise of VW Hyi, the system first reddens before becoming bluer and bluer. A comparison of both sets of observations during decay is unfortunately impossible since he did not provide the optical magnitudes during decay in his figures 1 and 2.

It therefore appears that (i) both systems are, for a given optical magnitude, bluer during decay than during rise by about 0.2 to 0.4 mag, and (ii) that the decay occurs at a roughly constant $B-V$. SS Cyg becomes bluer during the rise, while VW Hyi initially reddens before evolving in a similar way as SS Cyg. These are the main characteristics that, to be successful, the DIM should be able to reproduce.

\section{Model \label{sec:model}}
In order to model the light curve in various bands, we include the contributions from the accretion disc, the secondary star, the hot spot and the white dwarf, following the method described in \citet{shl03} that we briefly summarise here. The contribution from the boundary layer, even in the $U$ band, is negligible as compared to the disc and white dwarf contribution, and, although it has been included, is not described here; we refer to \cite{shl03} for more details.

\subsection{Disc contribution}
We use the code described in \citet{hmd98} for the DIM. As we consider the hot spot to be a separate source of light, heating of the outer rim of the accretion disc as described in \citet{bhl01} is not taken into account. The inner disc may be truncated, either because the white dwarf is magnetised, in which case truncation occurs at the magnetospheric radius or because of a transition to an optically thin flow occurring close to the white dwarf. In the latter case, we assume the same dependence of the inner disc radius as a function of the mass accretion rate as in the magnetic case, thereby introducing an effective magnetic dipole moment $\mu$. Indeed, observations indicate that, during quiescence, the accretion disc does not reach the white dwarf surface \citep[see e.g.][for recent references]{b15,bw19}. We calculate the time-dependent effective temperature as a function of the radius, and we then generate disc spectra by summing ATLAS9 Kurucz ODFNEW/NOVER stellar atmosphere models \citep{cgk97} from each disc ring. This library covers the range $3500 {\rm K} \leq T_{\rm eff} \leq 50\,000$K, and $0 \leq \log g \leq 5$, $T_{\rm eff}$ being the effective temperature and $g$ the surface gravity. We interpolate $F_\lambda / F_{\lambda , BB}$ where $F_\lambda$ and $F_{\lambda, BB}$ are the flux provided by the library and the blackbody flux respectively. When $T_{\rm eff}$ or $\log g$ falls outside the tabulated values, we use the last available value of $F_\lambda / F_{\lambda , BB}$. We then estimate the $UBVR$ magnitudes using the tabulated transmissions by \citet{b90}.

This way of calculating spectra neither takes into account the fact that in a disc gravity increases with altitude, nor the fact that energy is dissipated in the atmosphere. When calculating the vertical profiles and the relation between the local effective temperature and surface density, we do, however, take into account energy dissipation throughout the disc from the midplane up to the photosphere. One of the consequences of this approach is that a disc spectrum obtained by summing stellar spectra predicts strong Balmer jumps, that are either not observed or much weaker than predicted \citep{ld89}. Moreover, the strong emission lines observed during outbursts are also not predicted by the models; these lines indicate the presence of optically thin emitting material, possibly in form of a wind \citep[see e.g.][]{mkl15}. But, given the fact that the energy dissipation in optically thin layers, despite recent progresses, is not well constrained, we refrain from using a more sophisticated treatment, as done e.g. by \citet{wh98}. We instead consider the two extreme cases of blackbodies and stellar spectra. 

\subsection{Irradiated secondary}

The secondary is irradiated by the accretion luminosity; the irradiated side reaches a temperature given by:
\begin{equation}
\sigma T_{2,\rm irr}^4 = \sigma T_2^4 + (1-\eta ) \frac{1}{2} \frac{G M_1 \dot{M}}{r_{\rm wd}} \frac{1}{4 \pi a^2}
\end{equation}
where $\sigma$ is the Stephan-Boltzman constant, $T_2$ the unirradiated temperature of the secondary, $\eta$ the albedo, $M_1$ the white dwarf mass, $r_{\rm wd}$ its radius, $\dot{M}$ the accretion rate and $a$ the orbital separation. The factor 1/2 accounts for the fact that half of the accretion luminosity is emitted by the boundary layer in the EUV domain that is not able to penetrate deep in the atmosphere of the secondary and is thus inefficient for raising its temperature. 

The contribution of the secondary is then taken to be the average of the illuminated and unilluminated spectra, that we model as blackbodies. Given the oversimplified treatment of the variation of the effective temperature over the stellar surface, using accurate stellar spectra is not required.

\subsection{White dwarf}
The white dwarf spectrum is taken from \citet{ldc17} who provides theoretical spectra for hot DA white dwarfs with effective temperatures in the range 17\,000K -- 100\,000~K, and $7 \leq \log g \leq 9.5$. We also assume that the white-dwarf temperature remains constant during the outburst cycle. One expects that the white dwarf should be hotter at the end of an outburst than at its onset. \citet{gss17} reported a decrease of the white dwarf temperature in U Gem from 41\,500~K to 36\,250~K after a long (2 weeks) outburst. This effect is observed and expected \citep{pab05} to be significant only for long duration outbursts, and we do not take it into account here.

\subsection{Hot spot}
The hot spot luminosity is taken as \citep{s02}
\begin{equation}
L_{\rm hs} = \frac{1}{2} \frac{G(M_1+M_2)\dot{M_{\rm tr}}}{a} \Delta v^2
\end{equation}
where $\dot{M_{\rm tr}}$ is the mass transfer rate from the secondary, $M_2$ is the secondary mass, and $\Delta v^2$ is a dimensionless quantity of order of unity that depends on the degree of the disc's Roche-lobe filling, and, more weakly, on the mass ratio. A fit to $\Delta v^2$ is given by \citet{s02}. We ignore the possibility that the stream overflows the accretion disc. Stream overflow slightly modifies the outburst cycle, but this is a limited effect that can easily be compensated for by changes in the mass transfer rate or in the viscosity parameter. More importantly, it changes the luminosity due to the incorporation of matter at smaller radii as well as the color temperature of this contribution in a non trivial way; however, it is likely that both the luminosity and effective temperature would be higher than estimated here.

Observations indicate hot spot temperatures in the range $10^4 - 1.5 \times 10^4$~K \citep[see e.g.][and references therein]{bsg00}. For consistency with \cite{shl03}, we assume a blackbody spectrum with a temperature of 10$^4$~K, which is more appropriate for SS Cyg because of its long orbital period. This probably overestimates the contribution of the hot spot both in the B and V bands, but, as we shall see later, this contribution is by far not dominant and our results are relatively insensitive to the precise temperature of the hot spot.

\begin{table}
\caption{Model parameters}
\begin{center}
\begin{tabular}{lcc}
\hline\hline
System                                      &  SS Cyg    &  VW Hyi   \\
\hline
$P_{\rm orb}$ (h)                           &   6.60     &   1.78    \\
$M_1$ (M$_\odot$)                           &   1.00     &  0.70     \\
$M_2$ (M$_\odot$)                           &   0.67     &  0.10     \\
$\cos i$                                    &   0.7      &  0.7      \\
$\dot{M}_{\rm tr}$ (10$^{16}$ g~s$^{-1}$)   &   15       &  0.8      \\
$T_{\rm wd}$ (K)                            &  50\,000   &  20\,000  \\
$T_2$ (K)                                   &   4200     & 3200      \\
$\eta$                                      &    0.8     &   0.9     \\
$d$ (pc)                                    &  114       &   54      \\
$\alpha_{\rm c}$                            &   0.02     &   0.04    \\
$\alpha_{\rm h}$                            &   0.1      &   0.2     \\
\hline
\end{tabular}
\end{center}
\label{tab:param}
\end{table}

\section{Results \label{sec:results}}
\subsection{SS Cyg}

The parameters we use in the case of SS Cyg are given in Table \ref{tab:param}. The distance is from {\it Gaia} DR2 \citep{gaia,dr2}. We choose the primary mass $M_1=1$~M$\odot$ for the same reasons as discussed in \citet{hlk17}, in line with the recent determination by \citet{hsh17}, and take the mass ratio $q=M_2/M_1$ from \citet{bmr07}. From an analysis of the ellipsoidal variations, \citet{bmr07} limit the inclination to the range $45^\circ < i < 56^\circ$, in line with the absence of eclipses and the presence of orbital humps; we adopt here $i=45^\circ$. The effective temperature we use (4200~K) is less than the temperature corresponding to the observed spectral type of the secondary (K4V - K5V), but the secondary is highly spotted \citep{wnj02}, and this low temperature is needed to fit the observed visual magnitude during quiescence. One could have used a temperature of 4500~K corresponding to a K5 dwarf, with a surface reduced by 19\% due to the presence of dark spots that would have given the same $V$ magnitude for the secondary, with a $B-V$ reduced by 0.11 mag. Given the dilution of the secondary light, the change in $B-V$ during quiescence would be only 0.05 mag; both options therefore give essentially identical results. One should also be aware that, the luminosity of the secondary deduced by Bitner et al. (2007)  is compatible only with a distance larger that 140 pc (R. Robinson, private communication), which we now know to be contrary to observations. The white dwarf temperature is taken from \citet{sgm10}.

\begin{figure}
\includegraphics[width=\columnwidth]{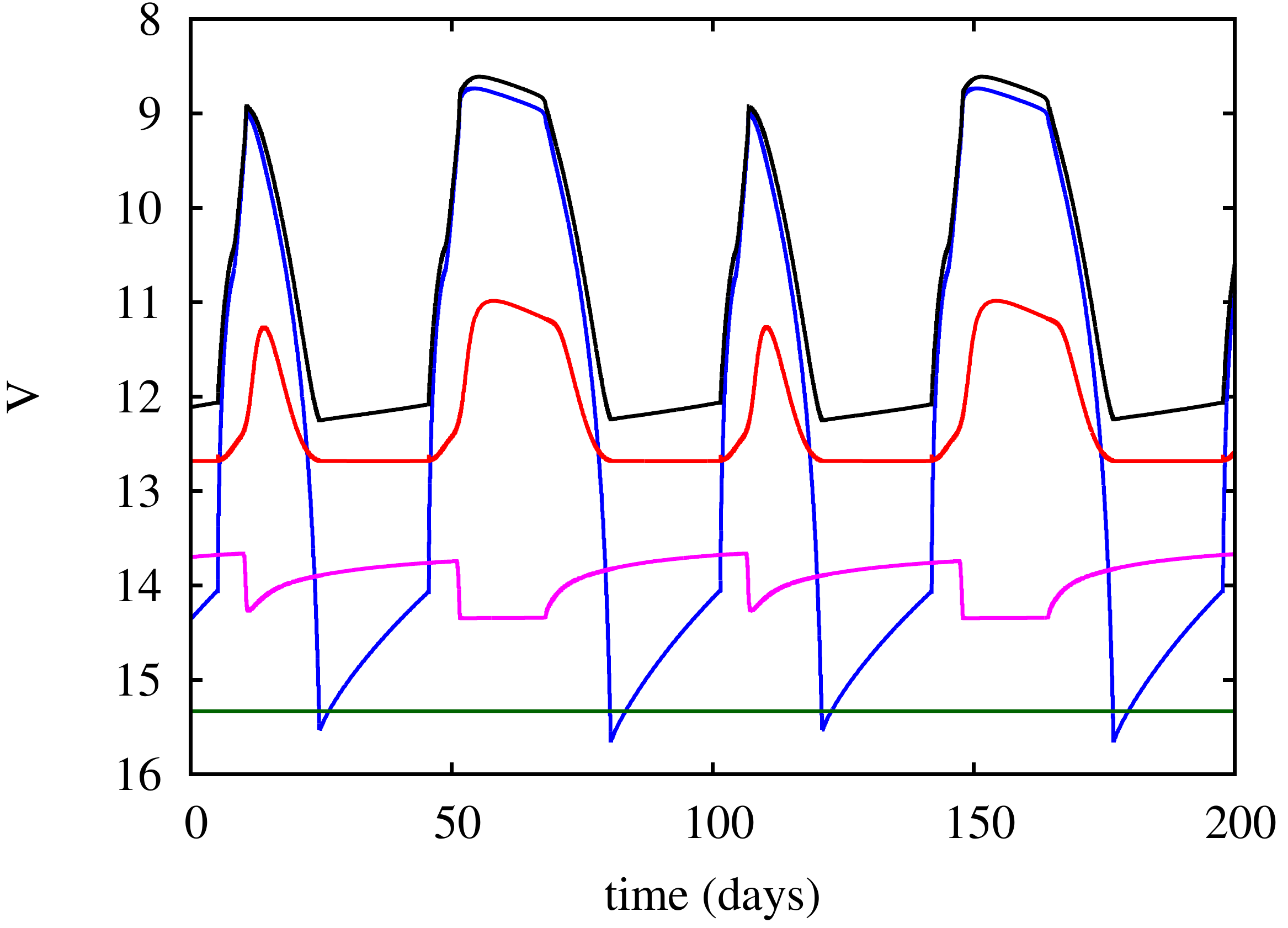}
\caption{Luminosity variations in the $V$ band of each component in SS Cyg in the case where the inner disc is not truncated. The blue curve represents the accretion disc, the red one the contribution from the secondary, the magenta one that from the hot spot, the green one the constant light from the white dwarf. The total optical luminosity is shown in black.}
\label{fig:sscyg_lc}
\end{figure}

\begin{figure}
\includegraphics[width=\columnwidth]{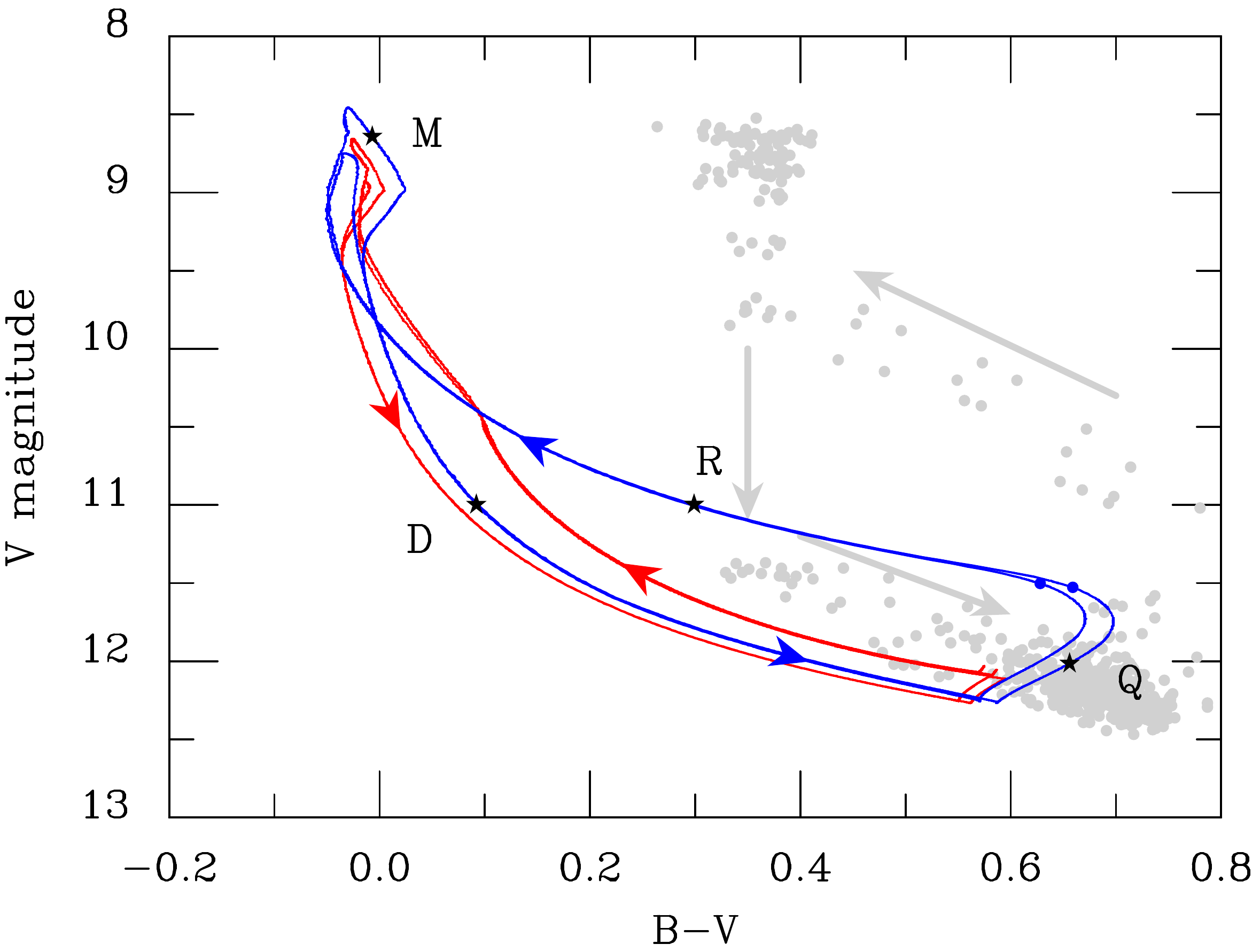}
\caption{Colour variations predicted by the model for parameters corresponding to SS Cyg. The blue curve shows the case where the inner disc is truncated, the red one when it extends down to the white dwarf surface. For the sake of clarity, $B-V$ has been increased by 0.02 mag in the truncated case, so that the blue and red curves do not superimpose during decay. The arrows indicate the direction of the time evolution. The dots at $V \sim 11.5$ indicate the onset of the outburst in the truncated case. The differences between the short and long outbursts are visible, but small. The star symbols denote the positions at which spectra are shown in Fig. \ref{fig:spectra}, and the grey dots are the observational data points.}
\label{fig:sscyg_cv}
\end{figure}

\begin{figure}
\includegraphics[width=\columnwidth]{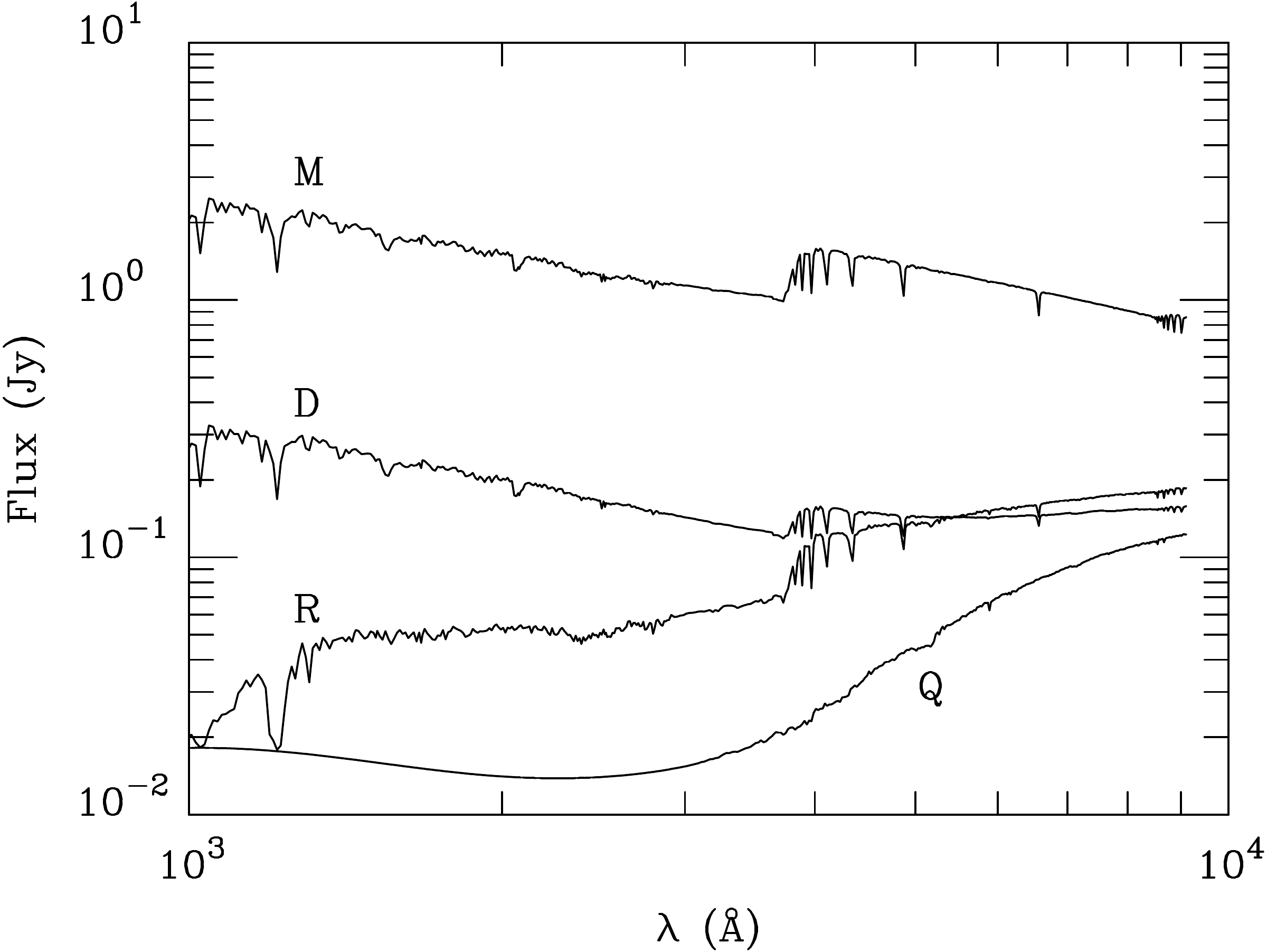}
\caption{Spectral evolution during an outburst of SS Cyg: quiescence (Q), rise (R), plateau (M) and decay (D). The corresponding locations in the colour-magnitude diagram are indicated in Fig. \ref{fig:sscyg_cv}.}
\label{fig:spectra}
\end{figure}

\begin{figure}
\includegraphics[width=\columnwidth]{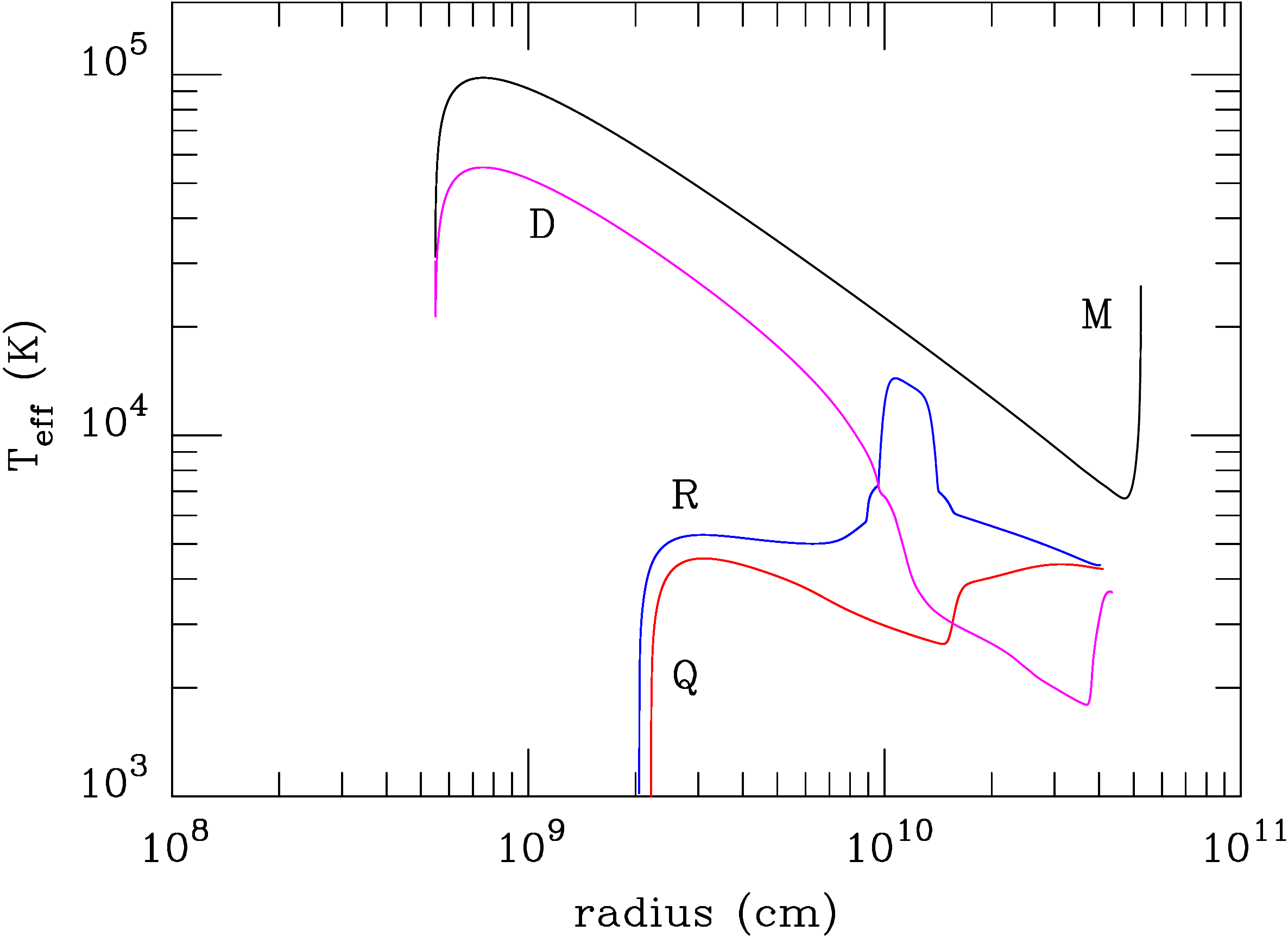}
\caption{Radial profiles of the disc effective temperature at the four epochs shown in Fig. \ref{fig:spectra}.}
\label{fig:profile}
\end{figure}

\begin{figure}
\includegraphics[width=\columnwidth]{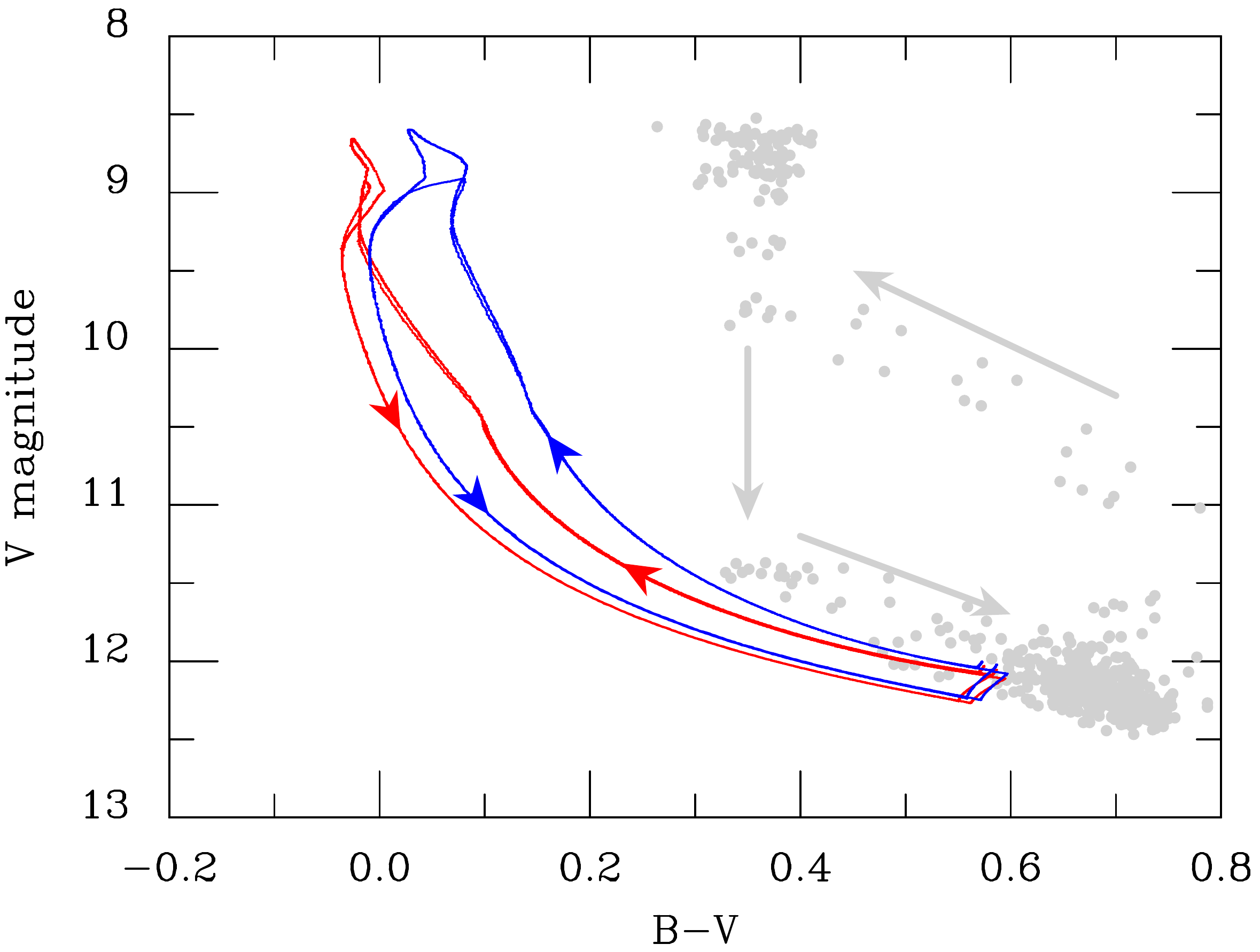}
\caption{Same as Fig.\ref{fig:sscyg_cv}, when the disc spectrum has been calculated using stellar spectra (red curve), or blackbodies (blue curve). In both cases, the disc is not truncated.}
\label{fig:sscyg_cv_kur}
\end{figure}

\begin{figure}
\includegraphics[width=\columnwidth]{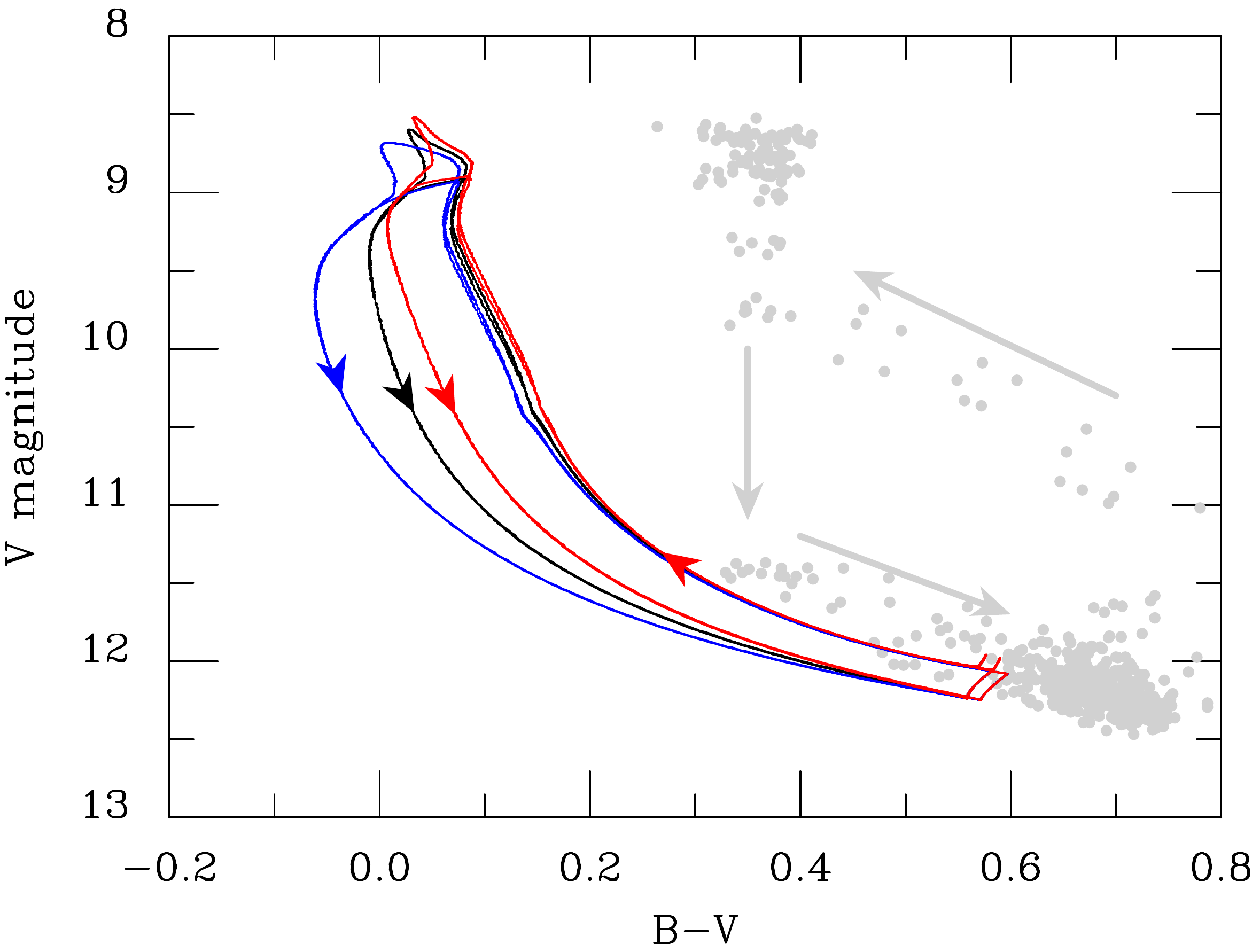}
\caption{Same as Fig.\ref{fig:sscyg_cv}, for different values of the albedo: 1.0 (blue curve), 0.8 (black curve) and 0.6 (red curve). In all three cases, the disc in not truncated and the spectra are calculated using blackbodies.}
\label{fig:sscyg_cv_irr}
\end{figure}

We use $\alpha_{\rm h}=0.1$ on the hot branch, and $\alpha_{\rm c}=0.02$ in the cold state. The mass transfer rate was set to $2 \times 10^{17}$ g~s$^{-1}$, so that the average recurrence time is 48 days when the disc is not truncated. When the disc extends to the white dwarf surface, the light curve, shown in Fig. \ref{fig:sscyg_lc} consists of a sequence of alternating long (about 30 days) and short (15 days) outbursts, during which respectively 41\% and 9\% of the total disc mass is accreted. These outbursts are significantly longer than those observed, but, as we shall see, the colour evolution is about the same in both cases. We also consider the case where the disc is truncated; we use a magnetic moment $\mu = 2 \times 10^{30}$ G~cm$^3$, corresponding to an inner radius of $2.2 \times 10^9$~cm in quiescence, about 4 times the white dwarf radius. The outbursts are also bimodal and have similar properties, except that the recurrence time is twice as long.

Figure \ref{fig:sscyg_lc} also shows the contribution of each component of the system to the total light curve. During quiescence, most of the optical light originates from the secondary, with some contribution from the hot spot. During outbursts, the disc dominates, with a small contribution from the irradiated secondary; the decrease of the hot spot luminosity during outbursts results from the disc expansion. 

Figure \ref{fig:sscyg_cv} shows the colour variations predicted by the model. There is a clear hysteresis in both cases. We first consider the case where the disc is not truncated. At the end of an outburst, the $V$ magnitude is at minimum; during quiescence, the luminosity slowly raises, the system becomes redder because the disc effective temperature raises, but is comparable to the secondary temperature and much smaller than the hot spot temperature which provides a significant fraction of the luminosity in the $B$ band. At some point, an instability is triggered in the inner disc, and the effective temperatures becomes larger than 10$^4$~K, so that the system become bluer and brighter. When the system is close to its maximum, the colour remains roughly constant, with a (limited) variation pattern, that is rather complex. During decay, the system initially evolves at a constant $B-V$, while the disc is fully in the hot state, and then becomes bluer and fainter while a cooling front propagates, bringing the outer part of the disc to temperatures low enough that they do not contribute any longer to the total luminosity, while the innermost-disc luminosity and temperatures still keep increasing for a while; during the final decay, the disc temperatures decrease everywhere, and the system gets fainter and reddens until the disc is fully brought to the cool state and the luminosity is at its minimum.

The fundamental reason for the colour difference at a given magnitude between raise and decay is that the disc is less massive during decay than during rise, and that the mass distribution with radius is also quite different; when the eruption starts, the surface density $\Sigma$ approximately follows the critical surface density profile, and increases with radius, while the effective temperature is roughly constant. On the other hand, just after maximum, the disc is not far from steady state, and $\Sigma$ and $T_{\rm eff}$ typically decrease as $r^{-3/4}$. This is the same reason that explains the hysteresis of SS Cyg in a diagram combining optical, X-rays and EUV fluxes \citep{hlk17}.

The evolution of a disc truncated in quiescence is similar, except that the luminosity and colour variations in quiescence are more important than in the non-truncated case, because, as a result of truncation, the disc is allowed to be brighter and represents a larger faction of the total luminosity (in quiescence the accretion rate strongly increases with radius). The colour evolution in quiescence is initially the same as in the non-truncated case: the system reddens, until the effective temperature of the inner disc becomes comparable to that of the secondary (the maximum effective temperature for the disc to stay on the cold branch is 5800~K), at which point $B-V$ decreases. 

It is interesting to note that the evolution in the colour-magnitude diagram is almost the same for short and long outbursts. This occurs because the heat front is able to reach the outer disc edge in both cases, and matter redistribution in the disc is efficient even for short outbursts.

Figure \ref{fig:spectra} shows the spectral evolution of the modelled system during a large outburst when the disc is truncated. In quiescence, the red part of the spectrum is dominated by the secondary while the hot spot is responsible for the observed light in the blue and UV domain; the accretion disk is essentially not detectable. The spectra calculated during the rise and decay are very different, although the visual magnitude is the same (11.0) for both spectra; in both cases, the disc dominates, but during rise, the total spectrum resembles that observed during quiescence. This is because during rise, the mass distribution in the disc is very similar to that in quiescence, whereas during decay, the mass distribution, the effective temperature, and hence the disc spectrum, are close to that of a steady disc; the spectrum during decay is essentially the same as at maximum. This is well illustrated by Fig. \ref{fig:profile} that shows the radial dependence of the disc effective temperature. In quiescence, the effective temperature does not vary much with radius, and is of order of $3000 - 4000$~K. The spectrum labelled `R' corresponds to the case where two heating fronts are propagating in the disc; the portion $1.0 \times 10^{10}$~cm $< r < 1.4 \times 10^{10}$~cm is on the hot branch, whereas the temperatures of innermost and outermost portions of the disc are those of quiescence. At maximum, the disc is close to being steady; the rise in $T_{\rm eff}$ at the disc edge is due to tidal dissipation that is important because the disc has reached the tidal truncation radius. During decay, the outer portion of the disc is brought on the cold branch, which is cooler than during rise because mass has been transferred to the inner disc; these inner portions are still on the hot branch, with $T_{\rm eff}$ close to a steady profile.

The colour-magnitude diagram significantly depends on the assumptions made for calculating the disk spectrum. Figure \ref{fig:sscyg_cv_kur} compares the colour-magnitude diagrams obtained when assuming that the spectrum calculated at a given radius is a blackbody (blue curve), or is a stellar spectrum (red curve). Whereas the decay phase is relatively independent of the assumption made, the system is, for a given optical magnitude, redder when one assumes blackbody spectra. This occurs because the Balmer jump during decay is small, so that the local disc spectrum is, in the visible domain, close to that of a blackbody, whereas the jump is large during the rising part; when $V=10$, the Balmer jump is about 0.86 mag if the system rises, and 0.43 if it is declining. This is consistent with observations that show that the Balmer jump is stronger during rise than decline, but as was already noted by \citet{ld89}, the jump is stronger than observed, in particular during the declining part where the jump is observed to be almost zero. \citet{ld89} concluded that energy dissipation in the atmosphere, not accounted for in Kurucz stellar models, must be significant. 

Figure \ref{fig:sscyg_cv_irr} illustrates the influence of the secondary irradiation on the colour-magnitude diagram. Irradiation produces additional light that reddens the spectrum emitted by the disc and white dwarf alone; we considered 3 possible values for the albedo: 1.0 (no irradiation), 0.8 (our reference value), and 0.6 (large irradiation effect). As can be seen, $B-V$ increases with decreasing albedo, as expected. One also notes that the rise is not much affected by irradiation, whereas the effect is quite significant during the decay phase. This occurs because, for a given optical magnitude, the mass accretion rate onto the white dwarf is much higher during the decay than during the rise; as an example, for $V=10$, $\dot{M}$ is $5.6 \times 10^{16}$~g~s$^{-1}$ during rise and $3.8 \times 10^{18}$~g~s$^{-1}$ during decay. 

When compared to observed outbursts, one notes that the quiescence and outburst magnitudes agree well with observations, which should not come as a surprise because the model parameters were chosen to assure that. The colour during quiescence agrees also reasonably well with observations, as well as the general evolution scheme. The amplitude of the hysteresis (a difference of 0.2 mag at $V=11$) is not grossly different -- but significantly smaller -- from what is observed (0.4 mag). On the other hand, the system is too blue at outburst maximum by about 0.4 mag. This discrepancy probably has the same root cause as the smaller-than-observed amplitude of the hysteresis cycle. Given that at maximum more than 90\% of the optical light is emitted by the disc, and that the disc is close to being steady, the disc spectrum at maximum is not very sensitive to the model parameters. We, however, anticipate that explaining the observed behaviour in $U$ will not be straightforward, as the observed colour evolution is extremely different when considering $B-V$ and $U-B$, whereas one would have expected a relatively similar behaviour.

\subsection{VW Hyi}

\begin{figure}
\includegraphics[width=\columnwidth]{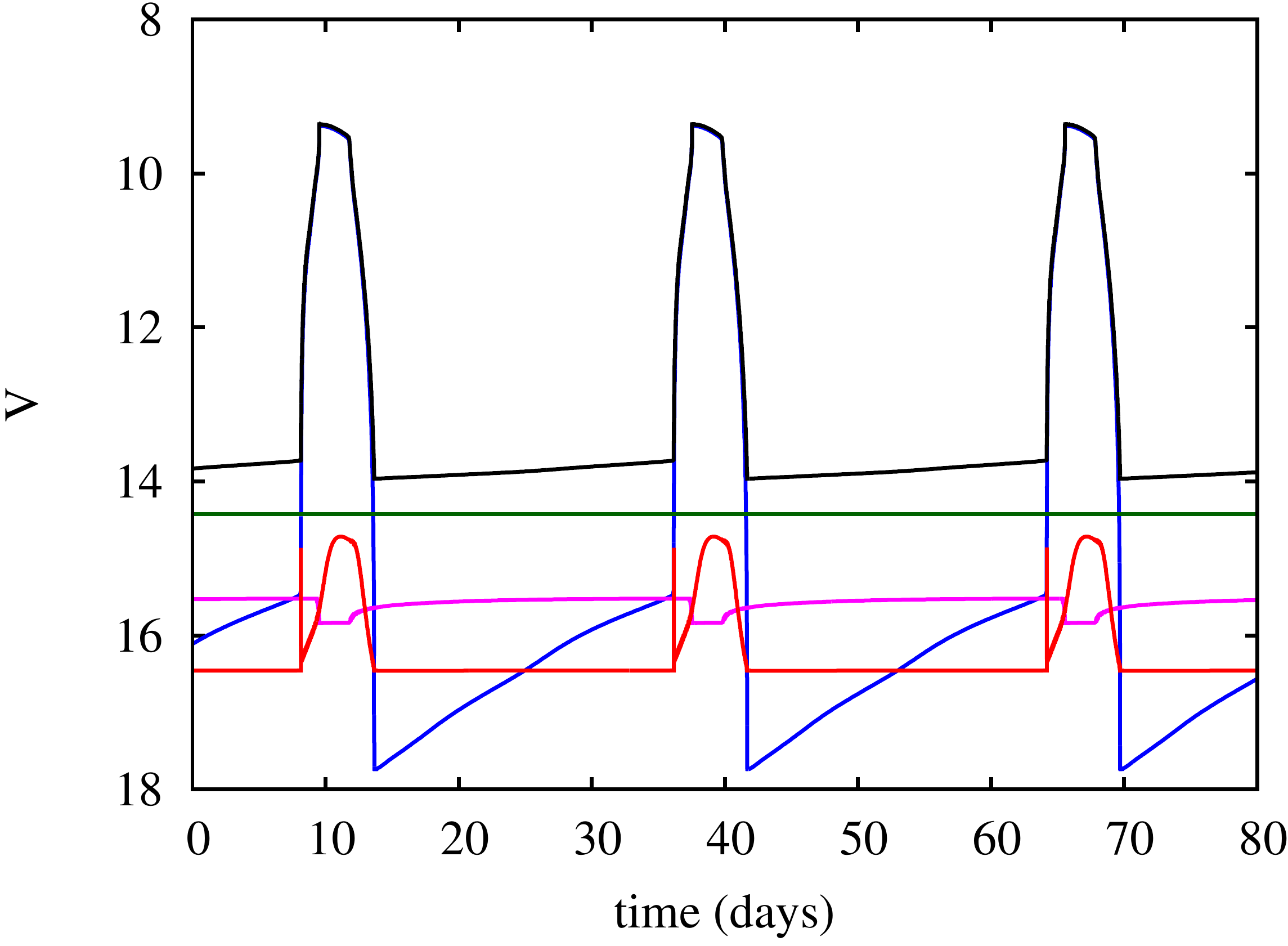}
\caption{Same as Fig. \ref{fig:sscyg_lc} for parameters corresponding to VW Hyi.}
\label{fig:vwhyi_lc}
\end{figure}

\begin{figure}
\includegraphics[width=\columnwidth]{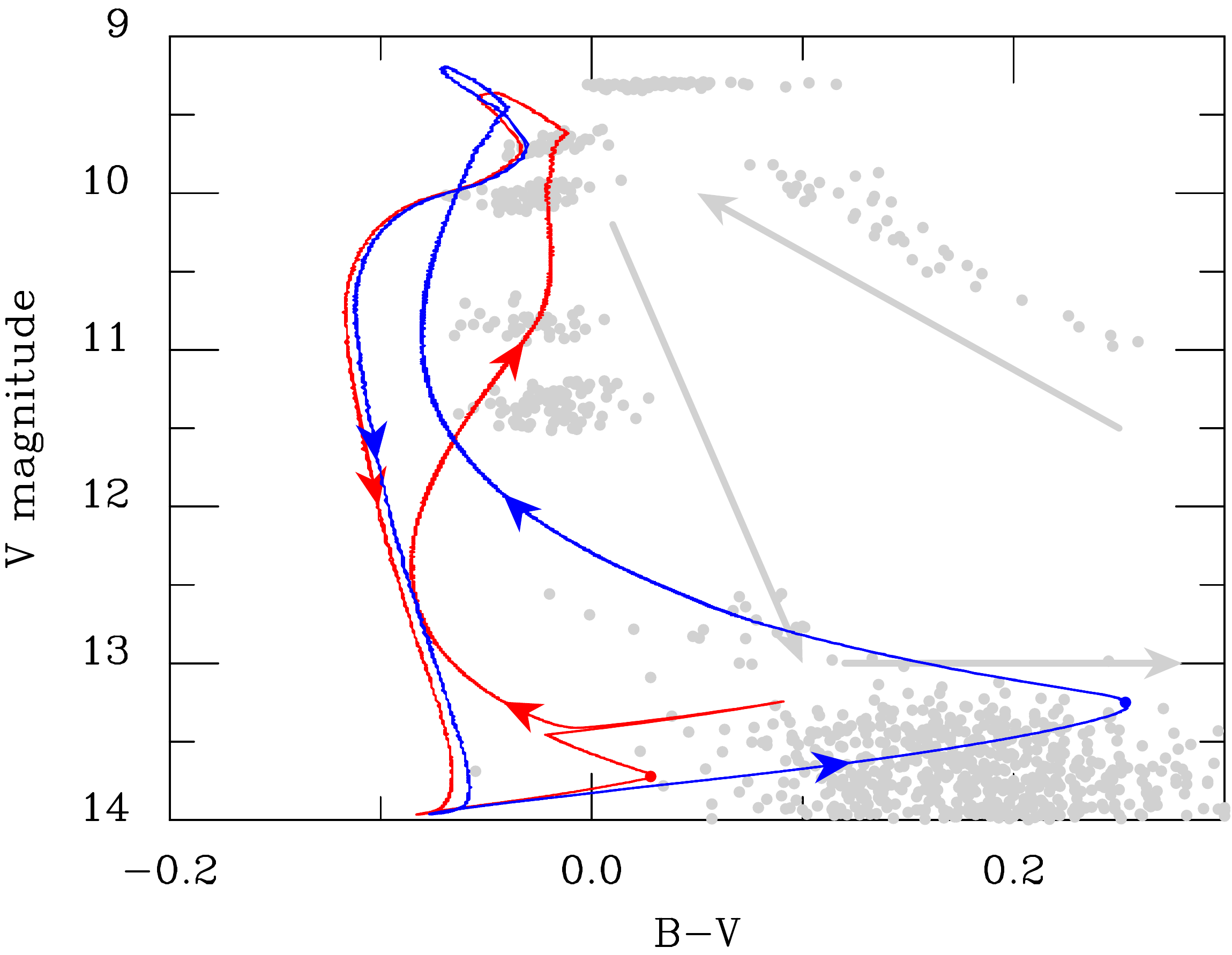}
\caption{Same as Fig. \ref{fig:sscyg_cv} for parameters corresponding to VW Hyi. In the case of disc truncation, the track followed by VW Hyi has been shifted to the left by 0.005 mag.}
\label{fig:vwhyi_cv}
\end{figure}

The parameters for VW Hyi are given in Table \ref{tab:param}. The distance is from {\it Gaia} DR2 \citep{gaia,dr2}. The primary and secondary masses are rather uncertain; \citet{shh06} quote a primary mass of $0.71^{+0.18}_{-0.26}$~M$_\odot$ based on radial velocities, while \citet{lgk09} determine a mass of $0.93$~M$_\odot$ based on spectral fitting of the white dwarf emission; the latter determination is based on a distance of 65 pc, and \citet{lgk09} noted it should be revised to lower masses for larger distances, with 0.71~M$_\odot$ for a distance of 74 pc. Given the {\it Gaia} distance, this would imply a mass close to 1~M$_\odot$. Here, we have chosen to use the spectroscopic determination of the masses, $M_1 = 0.7$~M$_\odot$. The secondary mass is then estimated from the mass ration $q = 0.147 \pm 0.05$ found by \citet{pkh05} from the superhump period. The white dwarf temperature is found to vary between 24\,000~K at the end of a superoutburst to 20\,000~K  before the next outburst. We use here $T_2 = 20\,000$~K. The secondary spectral type is mid to late M  \citep{hht11}; we adopt here a secondary temperature of 3200~K.

We use $\alpha_{\rm h}= 0.2$ on the hot branch, and $\alpha_{\rm c}= 0.04$ in the cold state; these values are twice as high as those used for SS Cyg, but the ratio of the two is the same. With a mass transfer rate of $8 \times 10^{15}$ g~s$^{-1}$, we obtain outbursts lasting for 5~d and recurring every 28~d when the disc is not truncated. When the disc is truncated with an inner disc radius of $1.8 \times 10^9$~cm during quiescence, that is 2.3 times the white dwarf radius ($\mu = 10^{30}$~G~cm$^3$), these numbers are 6~d and 50~d respectively. The light curve is shown in Fig. \ref{fig:vwhyi_lc}, together with the contributions from the various components of the system. There are significant differences with SS Cyg. First, the smaller disc size results in shorter outbursts cycles, both in terms of outburst duration and recurrence times; the latter varies as the square root of the outer disk radius \citep{l01}, while the outburst duration depends both on the disc size and the mass transfer rate. In addition, the quiescence luminosity is dominated by the white dwarf that represents almost 2/3 of the visual luminosity, with a significant contribution from the hot spot, and, to a lesser extent, from the secondary and the disc that becomes the second contributor just prior to an outburst. One can also note a sharp increase in the secondary luminosity at the onset of an outburst, that is caused by a sharp peak in the accretion rate occurring when the inner heating front reaches the white dwarf surface. This peak, lasting for less than a minute, is not physical.

Figure \ref{fig:vwhyi_cv} shows the corresponding colour-magnitude diagram. The main differences with SS Cyg is that VW Hyi is much bluer, because the secondary's contribution is much smaller. The spike that appears at $V \sim 13.5$ in the non-truncated case is due to the peak in the accretion rate when the heating front reaches the inner disc edge.

As in the case of SS Cyg, the predicted magnitudes agree well with observations; the amplitude of the colour difference between rise and decay is slightly smaller than what is observed even though the decay is slightly bluer (typically 0.1--0.2 mag) than observed; as mentioned above, this is presumably due to our inability to model accurately the disc spectrum, even when the disc is steady. However, the model does not reproduce the reddening observed during the initial rise (the system is expected to redden during quiescence, but as soon as it enters into outburst, it becomes bluer). \citet{ld94} suggested that this initial reddening occurs because the outburst is initiated in the outer disc and that, even on the hot branch, the outer disc is too cool to contribute significantly in the UV and, to a lesser extent, in the blue. For a system such as VW Hyi in which the main contributor to the $B$ and $V$ bands is a hot white dwarf, this could in principle be a possibility; the effective temperature is of the order of $1.2 \times 10^4 r_{10}^{-3/4}\dot{M}_{17}^{1/4}$~K for the parameters of VW Hyi, $r_{10}$ being the radius in units of $10^{10}$~cm and $\dot{M}_{17}$ the local accretion rate in units of $10^{17}$ g~s$^{-1}$, so that the additional contribution of a ring entering the hot state would be redder than the white dwarf. The main problem with such an interpretation  is that we were not able to find reasonable parameters for which outbursts of the outside-in type would be triggered, unless it would be due to mass-transfer enhancement not giving rise to a superoutburst which would require some fine-tuning of the parameters. 

\section{Conclusion}
Observations show that dwarf novae follow a characteristic loop in the $(B-V,V)$ colour-magnitude diagram. Its overall properties can be accounted for by the DIM, with some deficiencies, however.

The agreement with observations during quiescence is very good, which might come as a surprise since modelling the disc spectra is notoriously difficult at low accretion rates. This is the result of the disc being extremely faint in quiescence for the two systems that we consider here: the main contributor to the optical light is the secondary star in the case of SS Cyg, and the primary for VW Hyi.

The evolution of a system in the colour-magnitude diagram is affected by the spectral modelling of the accretion disc for the raising part of the outburst, and by the magnitude of irradiation during the decay. Disc spectra are not well approximated by a summation of stellar spectra, because of energy dissipation in the disc atmosphere. As mentioned earlier, attempts to take this effect into account used simple prescriptions for the local dissipation rate \citep[e.g.][]{sw86,kh86,wh98} that might not be a good description of reality. In addition, our synthetic spectra show absorption lines, and the observed emission lines that must be produces in hot, optically thin regions, cannot be reproduced by our model -- nor by any of the above mentioned disc models. In principle, the coupling of MRI simulations with a radiative transfer model should enable one to calculate properly the spectra of accretion discs; however, MRI models still have difficulties in reproducing the values of $\alpha$ needed to account for the general timing properties of dwarf novae, making such a development of little interest for the time being.

We did not consider here the $U$ band because the effect of the Balmer jump will be even stronger than in the $B$ band. We calculated the predicted $(U-B,V)$ colour magnitude diagrams; they are very similar to the $(B-V,V)$ diagrams, with $U-B = 0.4$ in quiescence and -0.6 in outburst in the case of SS Cyg. This is quite different from the values given by \citet{b80} who finds that $B-V$ is identical in quiescence and at maximum. This is certainly due in part to our poor modelling of the disc spectrum, but also perhaps to the presence of hot and optically thin gas that radiates in the UV. 

In summary, the DIM can explain:
\begin{itemize}
\item The existence and orientation of this hysteresis, which is a direct consequence of the total mass variation of the accretion disc as well as of the differences in the mass radial distribution in the accretion disc between raise and decay. 
\item the approximate colour and magnitude ranges covered during the evolution.
\end{itemize}
On the other hand, the DIM cannot explain the "redder than observed" colours near the peak of the outbursts, and the observed U-band behaviour; these are the same problems that have plagued spectral discs models for long, and that reflect our inability to model even steady-state discs correctly.

\begin{acknowledgements}
We acknowledge with thanks the variable star observations from the AAVSO International Database contributed by observers worldwide and used in this research. This work was supported by a National Science Centre, Poland grant 2015/19/B/ST9/01099. JPL was supported by a grant from the French Space Agency CNES. This work has made use of data from the European Space Agency (ESA) mission {\it Gaia} (\url{https://www.cosmos.esa.int/gaia}), processed by the {\it Gaia} Data Processing and Analysis Consortium (DPAC, \url{https://www.cosmos.esa.int/web/gaia/dpac/consortium}). Funding for the DPAC has been provided by national institutions, in particular the institutions participating in the {\it Gaia} Multilateral Agreement.
\end{acknowledgements}

\bibliographystyle{aa}
\bibliography{col_var}

\end{document}